\documentclass[11pt]{article}
\usepackage{epsfig}
\usepackage{latexsym}
\begin{document}
\thispagestyle{empty}
\begin{flushright}
hep-lat/0108001
\end{flushright}
\begin{center}
\vspace*{4mm}
{\huge Comparing lattice Dirac operators in 
\vskip2mm
smooth instanton backgrounds}
\vskip11mm
{\bf Christof Gattringer${\,}^{a),\dagger}$, Meinulf G\"ockeler${\,}^{a)}$, 
C.B.~Lang${\,}^{b)}$,
P.E.L.~Rakow${\,}^{a)}$ and  
Andreas Sch\"afer${\,}^{a)}$}
\vskip2mm
${}^{a)}$ Institut f\"ur Theoretische Physik \\
Universit\"at Regensburg \\
93040 Regensburg, Germany
\vskip3mm
${}^{b)}$ Institut f\"ur Theoretische Physik \\
Karl-Franzens-Universit\"at Graz \\
8010 Graz, Austria
\vskip20mm
\begin{abstract}
We compare the behavior of different lattice Dirac operators 
in gauge backgrounds which are 
lattice discretizations of a classical instanton.
In particular we analyze the standard Wilson operator,
 a chirally improved Dirac operator and
 the overlap operators constructed from these two 
 operators. We discuss the flow of real eigenvalues 
as a function of the instanton size. An analysis of the eigenvectors
shows that overlap fermions with the Wilson operator as input operator 
have difficulties with reproducing the continuum zero mode already for
moderately small instantons. This problem is greatly reduced 
when using the chirally improved operator for the overlap projection.
\end{abstract}
\end{center}
\vskip5mm \nopagebreak \begin{flushleft} \rule{2 in}{0.03cm}
\\ {\footnotesize \ 
${}^\dagger$ Supported by the Austrian Academy of Sciences (APART 654).}
\end{flushleft}
\newpage
\setcounter{page}{1}
\noindent
{\bf Introductory remarks:}
Recently it was realized that the Ginsparg-Wilson equation 
\cite{GiWi82} is crucial for implementing chiral symmetry on the lattice. 
Currently three types of exact solutions are known: The overlap operator
\cite{overlap}, perfect actions~\cite{fixpd} and domain wall 
fermions~\cite{domainwall}. Furthermore a systematic 
expansion of a solution of the Ginsparg-Wilson equation was developed and 
tested in~\cite{Gaetal00,Gaetal00b}.
These new lattice Dirac operators based on the Ginsparg-Wilson equation
were recently used to 
analyze relevant excitations of the QCD vacuum which affect the Dirac 
operator~\cite{degrand,edwards,Gaetal01,domainwall2}, in particular the
local chirality variable proposed in~\cite{locchir} was studied in detail.

These studies of the excitations affecting the lattice
Dirac operator are motivated by the instanton picture of chiral symmetry 
breaking (see~\cite{SchSh98} for extended reviews). A single instanton
or anti-instanton produces a zero eigenvalue of the Dirac operator. An 
interacting pair of an instanton and an anti-instanton 
leads to a complex conjugate pair of small
eigenvalues instead of two zero eigenvalues. 
In instanton models the QCD vacuum is pictured as a fluid 
of interacting instantons and anti-instantons which lead to an accumulation of
small eigenvalues near the origin. Since the density of eigenvalues
at the origin is related to the chiral condensate through the
Banks-Casher relation~\cite{BaCa80} the fluid of instantons and anti-instantons
leads to a breaking of chiral symmetry. The lattice studies
\cite{degrand}-\cite{domainwall2},\cite{wuppertal} 
tried to prove or refute this
picture of interacting instantons and anti-instantons.
To be more specific, various observables built
from the eigenvectors of the Dirac operator were studied for background
gauge fields generated by simulations in the quenched approximation.

A good method for testing properties of different lattice Dirac operators
is to study them in smooth instanton backgrounds.
In this letter we report on such a study comparing 
the overlap Dirac operator based on the Wilson operator~\cite{overlap}, 
the standard Wilson operator, a recently proposed approximate solution
of the Ginsparg-Wilson equation~\cite{Gaetal00,Gaetal00b} 
which we will refer to as
the {\sl chirally improved Dirac operator} and the overlap operator with
the chirally improved operator as input operator. We analyze different 
properties of the eigenvectors and eigenvalues 
for these Dirac operators using a lattice 
discretization of instantons.
Another study including smooth instanton backgrounds was
reported in~\cite{DeGrand}, which investigated the number of zero modes
and the topological charge as a function of instanton radius.

The goal of our analysis is twofold: Firstly it serves to better understand the
results of the
above mentioned studies of relevant excitations in the QCD vacuum as seen by
the lattice Dirac operator. When the instanton is large compared with
the lattice spacing, all operators give good results. But
when we consider smaller instanton radii we find that the 
overlap operator with Wilson input operator has difficulties with 
reproducing the continuum zero mode in an instanton 
background even for moderately small instantons. We show that this problem is 
greatly reduced 
when using the chirally improved operator for the overlap projection.

The second goal of this study is to achieve a better understanding
of technical aspects of the overlap projection. The
sensitivity to defects should be understood, in particular if one
uses an approximate solution of the Ginsparg-Wilson
equation as a starting point for the overlap projection as proposed in 
\cite{bietenholz}.
\\
\\
{\bf Technicalities:}
We analyze the overlap Dirac operator, the standard Wilson 
Dirac operator and the chirally improved Dirac operator. The latter has been 
described in detail in~\cite{Gaetal00}. Here,
since the discretized instanton configurations are smooth,
we use the coefficients for the free case as listed in the appendix 
of~\cite{Gaetal00b}. 

The overlap operator is given by
\begin{equation}
D_{ov} \; = \; 1 - \frac{1 - D_0}{\sqrt{(1 - D_0^\dagger)(1 - D_0)}} \; ,
\label{overlap}
\end{equation}
where $D_0$ is any reasonable lattice version of the Dirac operator.
Here we use for 
$D_0$ both the standard Wilson Dirac operator 
leading to the {\sl Wilson overlap operator} as well as the 
chirally improved operator leading to the {\sl chirally improved overlap operator}.
We construct the 
inverse square root using Chebychev approximation following the
approach discussed in~\cite{jansenetal1,jansenetal2}.

A method for putting an instanton on the lattice has been proposed in 
Ref.~\cite{clinst}. Here we use a slightly modified version of this 
procedure. In the regular gauge, the continuum instanton potential is
written as
\begin{equation}
 A_\mu (x) = \frac{\mathrm i}{2(x^2 + \rho^2)} 
  \left( s_\mu \overline{s}_\nu - s_\nu \overline{s}_\mu \right) x_\nu \,,
\end{equation} 
and in the singular gauge we have
\begin{equation}
 \tilde{A}_\mu (x) = \frac{\mathrm i \rho^2}{2 x^2(x^2 + \rho^2)}
  \left( \overline{s}_\mu s_\nu - \overline{s}_\nu s_\mu \right) x_\nu \,,
\end{equation} 
where $s_4 = \overline{s}_4 = 1$, 
$s_j = - \overline{s}_j = \mathrm i \sigma_j$ ($j=1,2,3$) and $\rho$ is the
``radius'' of the instanton. The corresponding expressions for an 
anti-instanton are obtained by exchanging $s_\lambda$ and 
$\overline{s}_\lambda$.

The first step of our procedure is a coordinate transformation 
$x_\mu \to y_\mu$ which maps the real line onto the interval $(0,L)$,
where $L$ is identified with the length of our lattice (measured in units
of the lattice spacing and assumed to be even). 
For this transformation we take 
$ x_\mu = f(y_\mu) $ with
\begin{equation}
 f(y) = L^2 \left[ (L-y)^{-1} - y^{-1} \right] \,.
\end{equation} 
Acting with this transformation on the (anti-)instanton potential 
either in the regular or in the singular gauge we obtain the 
corresponding potentials on the four-torus $(0,L)^4$. 
The center of the instanton, 
$x_1 = \cdots = x_4 = 0$, is mapped onto $y_1 = \cdots = y_4 = L/2$. 
Note that the instanton is ``squeezed'' by this procedure
so that the radius $R$ of the potential on the four-torus 
is related to the radius $\rho$ of its infinite-volume precursor
by $\rho = f(R+L/2)$.

In the second step we divide our lattice into an ``inner part'' around
$y_1 = \cdots = y_4 = L/2$ and a complementary ``outer part''. 
According to the procedure of Ref.~\cite{clinst} we work with the 
potential in the regular
(singular) gauge in the inner (outer) part using the gauge transformation
connecting the two gauges to glue both potentials together. 
The third step consists in computing the gauge links from the potential.
This is easily done analytically.

 The resulting SU(2) link variables are finally embedded in SU(3)
in the most trivial way, namely as $2 \times 2$ blocks in the upper 
left-hand corner of the SU(3) matrices.

We work on lattices with size $16^4$ and $12^4$. For the fermions we 
use periodic boundary conditions in space direction and anti-periodic 
boundary conditions in time direction.  The numerical computations
of eigenvalues and eigenvectors are done with the implicitly
restarted Arnoldi method~\cite{arnoldi}.
\\
\\
{\bf The flow of real eigenvalues:}
It is known that in an instanton background
an exact solution of the Ginsparg-Wilson equation 
has exact zero eigenvalues~\cite{fixpd}, as does the continuum Dirac operator. 
Currently only the overlap operator 
 shows this property. A practical implementation of perfect actions 
requires a finite parametrization of the lattice Dirac operator
and the coefficients of the parametrization are determined from renormalization
group transformations. Since an exact solution of the Ginsparg-Wilson equation
is necessarily non-ultralocal~\cite{horvath}, any practical implementation
of the perfect action will be an ultralocal approximation of a solution of
the Ginsparg-Wilson equation. Thus the fixed point operator as well as the
chirally improved operator will only have approximate zero modes. Both these
Dirac operators obey $\gamma_5$-hermiticity, 
i.e.~$\gamma_5 D \gamma_5 = D^\dagger$. This implies~\cite{itoh} that 
eigenvectors $\psi$ of $D$ with eigenvalues $\lambda$ have 
$\psi^\dagger \gamma_5 \psi = 0$ unless $\lambda$ is real. This  
has to be compared with the property that 
$\psi^\dagger \gamma_5 \psi = 0$ unless $\lambda$ is zero, which
holds for eigenvectors
of an exact solution of the Ginsparg-Wilson equation or for
the eigenmodes of the continuum Dirac operator. It thus follows that
for perfect actions, for the chirally improved Dirac operator and also 
for Wilson fermions  
only eigenvectors with real eigenvalues are possible candidates for 
topological modes.

\begin{figure}[t]
\begin{center}
\vspace*{3mm}
\hspace*{-10mm}
\epsfig{file=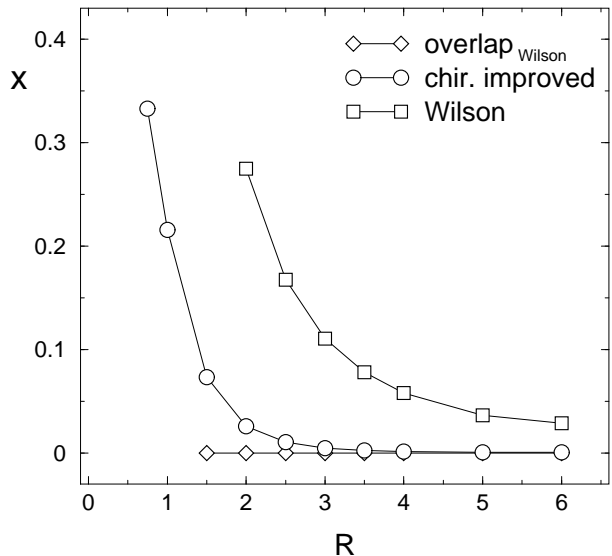,width=8cm,clip}
\caption{The dependence of the position $x$ of the
real eigenvalue (zero mode) 
on the radius $R$ (in lattice units)
of the underlying instanton. We show our results for
the Wilson overlap operator (diamonds), the chirally improved operator
(circles) and the Wilson operator (squares).
The data were computed on $16^4$ lattices.
\label{realdrift}}
\end{center}
\end{figure}

An interesting question is, how well different operators manage to 
project the real mode into the origin when the underlying gauge field changes. 
In Fig.~\ref{realdrift} we show the position $x$ of 
the real eigenvalue as a function 
of the radius $R$ (in lattice units)
of the underlying instanton configuration. This study was 
done on lattices of size $16^4$. We display data 
for the Wilson overlap operator, the chirally improved operator and the 
standard Wilson operator for identical gauge configurations. 
The behavior of the zero modes of the Wilson overlap operator is of course trivial
and it serves only as a reference line. For the Wilson operator we find a 
very strong dependence of the real mode on the radius of the instanton. 
Already for large instantons the corresponding eigenvalue is shifted 
to relatively large real values and this shift increases as the radius
of the instanton shrinks further. The chirally improved Dirac operator 
is considerably less sensitive to the radius of the instanton.
 It starts to deviate from 0 only for radii below 2.5 lattice units. 

Our analysis sheds light on a potential problem of the overlap projection:
Whenever the background configurations contain defects, 
i.e.~gauge configurations with very small excitations carrying topological 
charge, the overlap projection becomes numerically expensive and low
eigenvalues of $(1-D_0^\dagger)(1-D_0)$ have to be projected out
before the square root in (\ref{overlap}) can be evaluated 
numerically. The underlying
mechanism is nicely illustrated in our Fig.~\ref{realdrift}: For small
instanton radius the real eigenvalue of the Wilson operator $D_0$ used
in the overlap projection (\ref{overlap}) comes close to the center of the 
projection (1 in the complex plane) causing the inverse square root to blow
up which spoils the numerical evaluation of the overlap operator. Comparing
the Wilson curve with the curve for the chirally improved operator shows that 
when using already an approximate solution of the Ginsparg-Wilson equation
as $D_0$ in the overlap projection~\cite{bietenholz} the problem
with defects is milder. 
\begin{figure}[h]
\begin{center}
\vspace*{3mm}
\hspace*{-10mm}
\epsfig{file=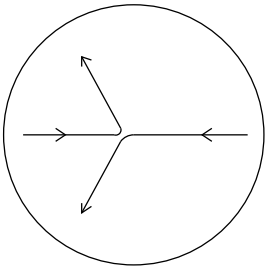,width=3.5cm,clip}
\caption{Schematic picture for the movement of the physical real mode 
and its doubler partner as a single instanton is destroyed. The circle
represents the Ginsparg-Wilson circle in the complex $\lambda$ plane, 
 and the two other curves are the schematic
trajectories of the two eigenvalues.
\label{sketch}}
\end{center}
\end{figure}

Fig.~\ref{realdrift} also illustrates how topological modes are treated 
by solutions and approximate solutions of the Ginsparg-Wilson equation. 
For an exact solution of the Ginsparg-Wilson equation it is known that the 
spectrum depends discontinuously on the underlying gauge field. Since 
(1) the total number of eigenvalues is even, (2) all eigenvalues have to 
lie on the Ginsparg-Wilson circle (the circle 
with radius 1 and center 1 in the complex plane)
and (3) all modes which are not real
come in complex conjugate pairs, a single eigenvalue 0
has to vanish discontinuously
as the underlying gauge field is deformed from topological sector 1
to sector 0. Since an ultralocal approximation of a solution of the 
Ginsparg-Wilson equation cannot show such discontinuous behavior 
the change of the
topological sector of the underlying gauge field has to manifest itself  
differently:
As the instanton shrinks, the real mode starts to travel into the interior
of the Ginsparg-Wilson circle (for an approximate solution of the 
Ginsparg-Wilson equation the eigenvalues are not confined to the circle)
where it meets a partner from the doubler branch of the spectrum. When 
they meet on the real axis they can continuously
form a complex conjugate pair and travel back to the outside of the circle. 
A schematic picture of this behavior is given in Fig.~\ref{sketch}.

\begin{figure}[t]
\begin{center}
\epsfig{file=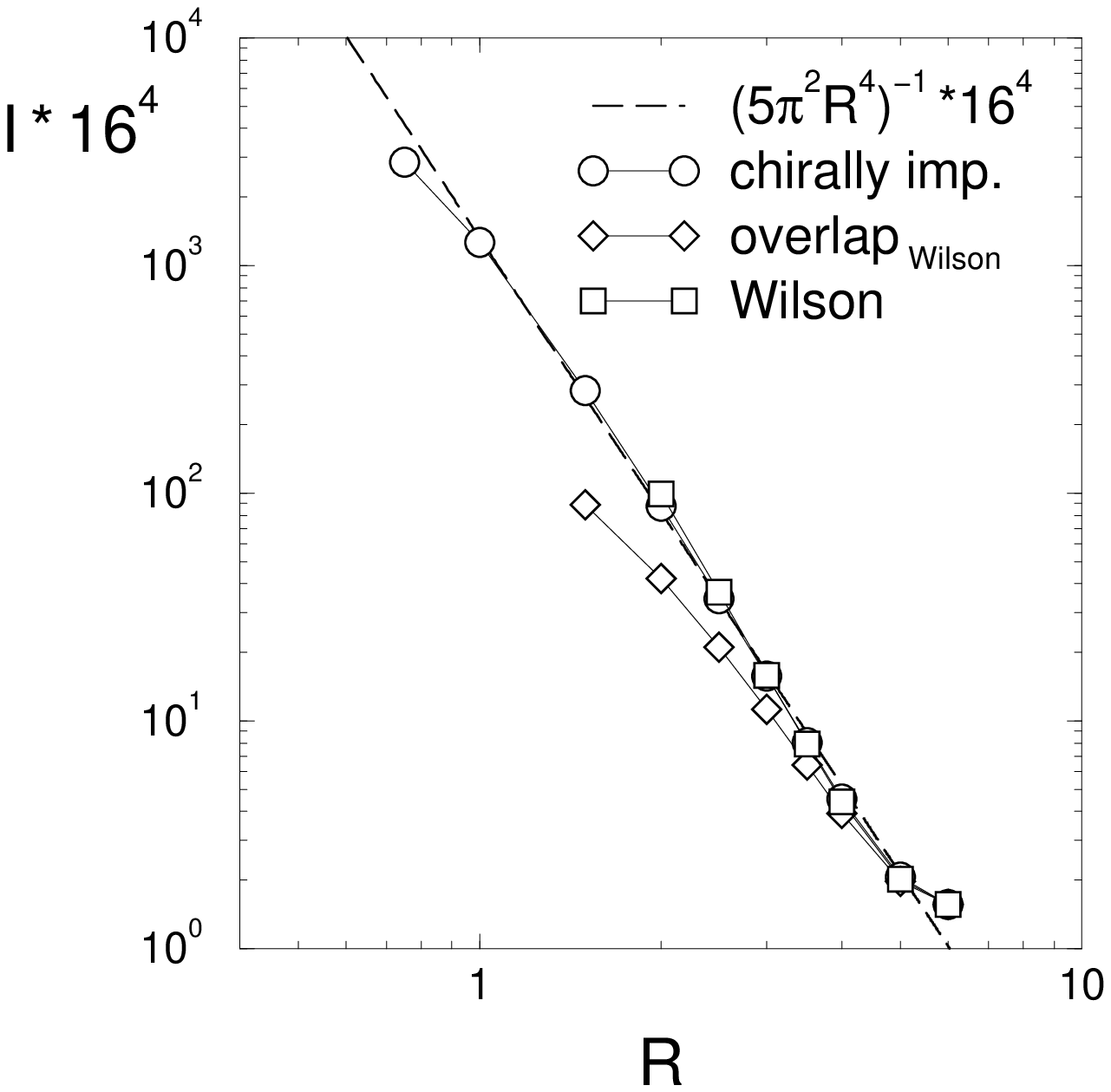,width=6.1cm}
\epsfig{file=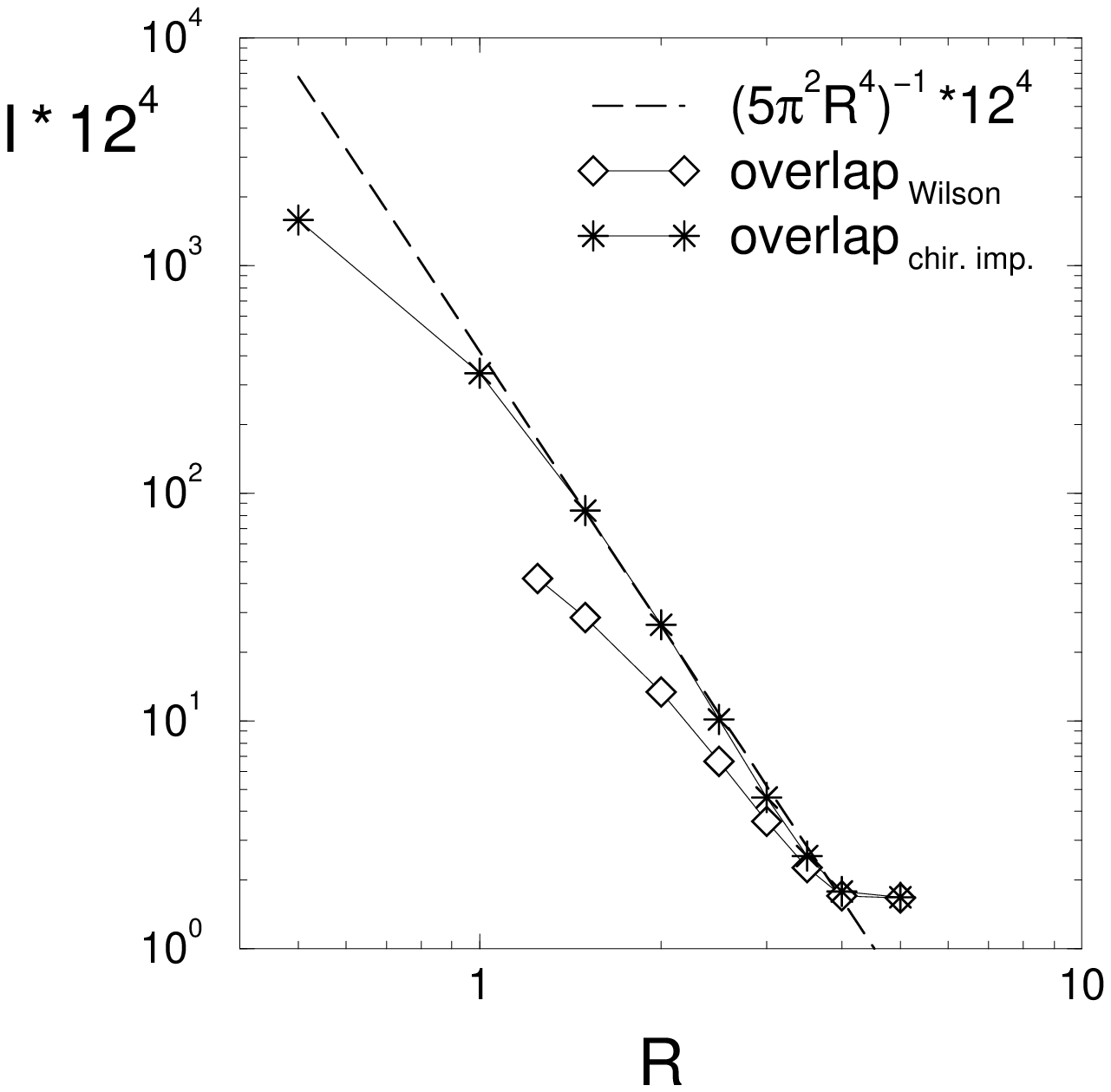,width=6.1cm}
\caption{The inverse participation ratio of the zero mode as a function of 
the radius $R$ of the underlying instanton. On the left-hand side 
we show our results on $16^4$ lattices for
the Wilson overlap operator (diamonds), 
the chirally improved operator
(circles) and the Wilson operator (squares). 
The right-hand side plot shows a comparison of the Wilson overlap 
operator (diamonds) and the chirally improved overlap operator
(asterisks) on $12^4$ lattices. The dashed lines represent the continuum formula.
We rescale $I$ by the volume, i.e.~we plot $I\times16^4$, respectively 
$I\times12^4$. 
\label{iprvsr}}
\end{center}
\end{figure}

The better a Dirac operator approximates a 
Ginsparg-Wilson fermion, the faster the eigenvalue moves through the center 
of the circle. We remark that we have also seen the behavior of 
 Fig.~\ref{sketch}, in which a 
partner from the doubler sector meets the displaced real mode to
form a complex conjugate pair, in a numerical study of random matrices
with the same symmetries as the lattice Dirac operator. 
\\
\\
{\bf Localization properties of the eigenvectors:} 
In order to further analyze the
behavior of different lattice Dirac operators in instanton backgrounds
we now study properties of their eigenvectors. In the background of an 
instanton field the continuum Dirac operator has a zero mode $\psi_0$
(see e.g.~\cite{SchSh98}).
It is localized at the same position as the underlying
instanton. A gauge invariant density $p(x)$ which inherits this
localization is obtained by summing over the color and Dirac indices
$c$ and $\alpha$,
\begin{equation}
p(x) \; = \; \sum_{\alpha,c} \psi_0(x)_{\alpha,c}^* \psi_0(x)_{\alpha,c} 
\; = \; 
\frac{2R^2}{\pi^2(R^2 + x^2)^3} \; .
\end{equation}
Due to the normalization of the zero modes we have $\int d^4x \, p(x) = 1$.
A measure of the localization of $\psi$ is
given by the inverse participation ratio, 
\begin{equation}
I \; = \; \int d^4x \, p(x)^2 \; = \; (5 \pi^2 R^4)^{-1}.
\label{iprclass}
\end{equation}
As an alternative measure of locality~\cite{DeGrand}
studies the self-correlation of the local chiral density.
For different radii of our lattice instantons we computed the 
inverse participation ratio of the corresponding zero mode. 
In the plot on the left-hand side of
Fig.~\ref{iprvsr} we show our results for lattice size
$16^4$. We normalized 
the inverse participation ratio $I$ by the volume\footnote{With this
normalization the inverse participation ratio is a quantity widely used in 
solid state physics.}, i.e.~we plot $I\times16^4$. The
symbols give the numerical results while the dashed line represents the
continuum formula from Eq.~(\ref{iprclass}).
From the plot is obvious that for our 
largest value of the instanton radius, $R=6$, the lattice results 
are slightly above the continuum value due to finite size effects. 
For smaller $R$, the numerical results for the two ultralocal 
lattice Dirac operators, i.e.~the Wilson Dirac operator (squares)
and the chirally improved Dirac operator (circles) follow the continuum
curve down to relatively small values of $R$. For the chirally improved 
operator the agreement with the continuum result holds down to 
$R=1$ and only for $R=0.75$, as the instanton begins to ``fall through the
lattice'' we find a considerable deviation. For the Wilson operator,
due to the move of the real mode into the interior of the eigenvalue
distribution we could not obtain data for $R < 2$. However, down to $R=2$ 
the Wilson operator also follows the continuum curve quite well. 

The situation is different for 
the Wilson overlap operator. Already at $R=3.5$ the zero mode of the 
Wilson overlap operator has a value of the inverse participation ratio 
which is visibly different from the continuum result, 
and the error quickly increases as $R$ is decreased (note the logarithmic
 scale). At $R = 2.5$ the Wilson overlap result amounts to 
only 60\% of the continuum formula and at $R = 1.5$ only about 30\% remain.

In the plot on the right-hand side of Fig.~\ref{iprvsr} we compare the inverse
participation ratio of the Wilson overlap operator (diamonds) to the inverse
participation ratio of the zero mode of the chirally improved overlap operator
(asterisks). It is obvious that 
with the chirally improved overlap operator the
results for small instantons are considerably closer to the 
continuum line.

\begin{figure}[t]
\begin{center}
\epsfig{file=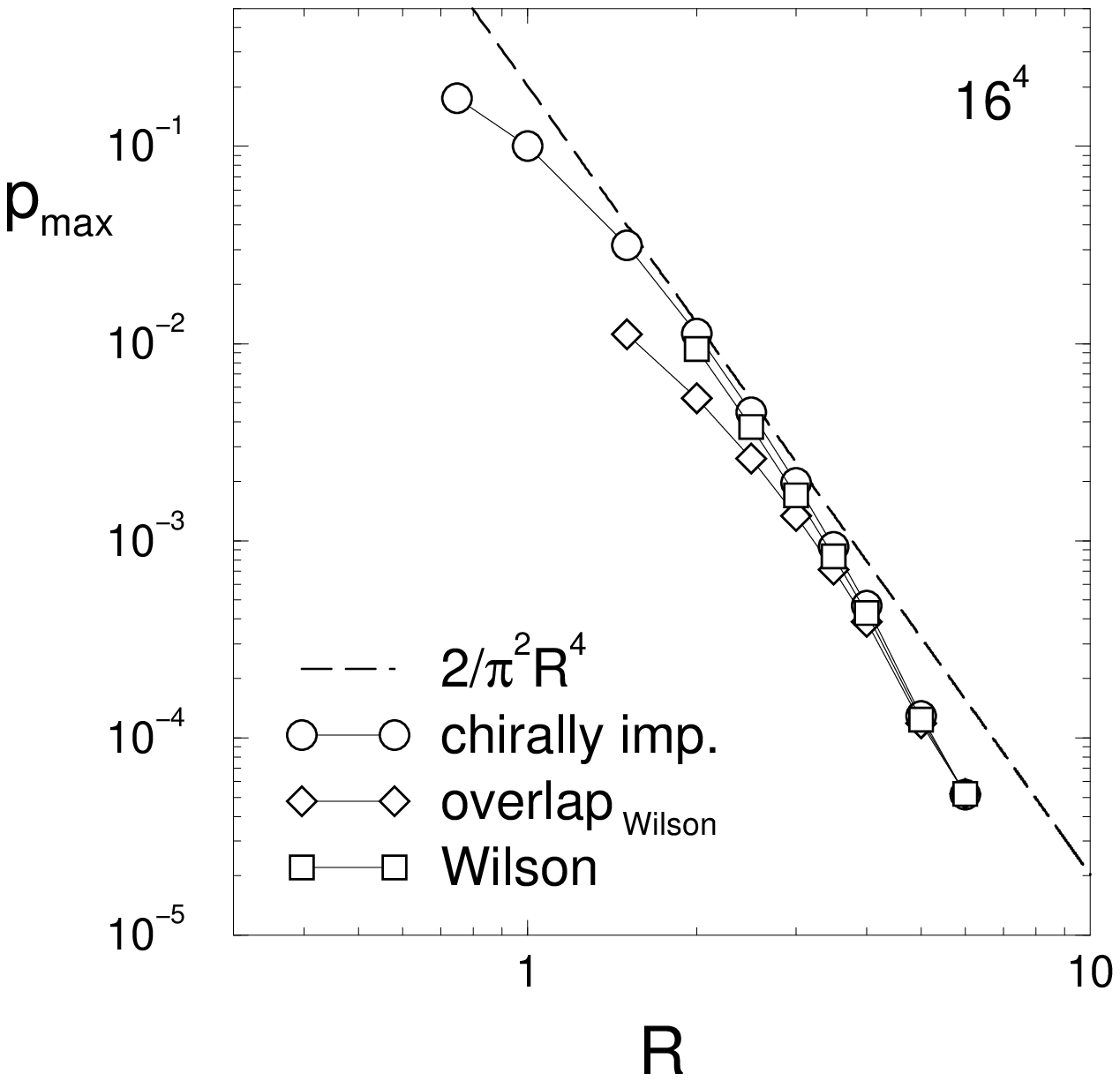,width=5.9cm}
\hspace{2mm}
\epsfig{file=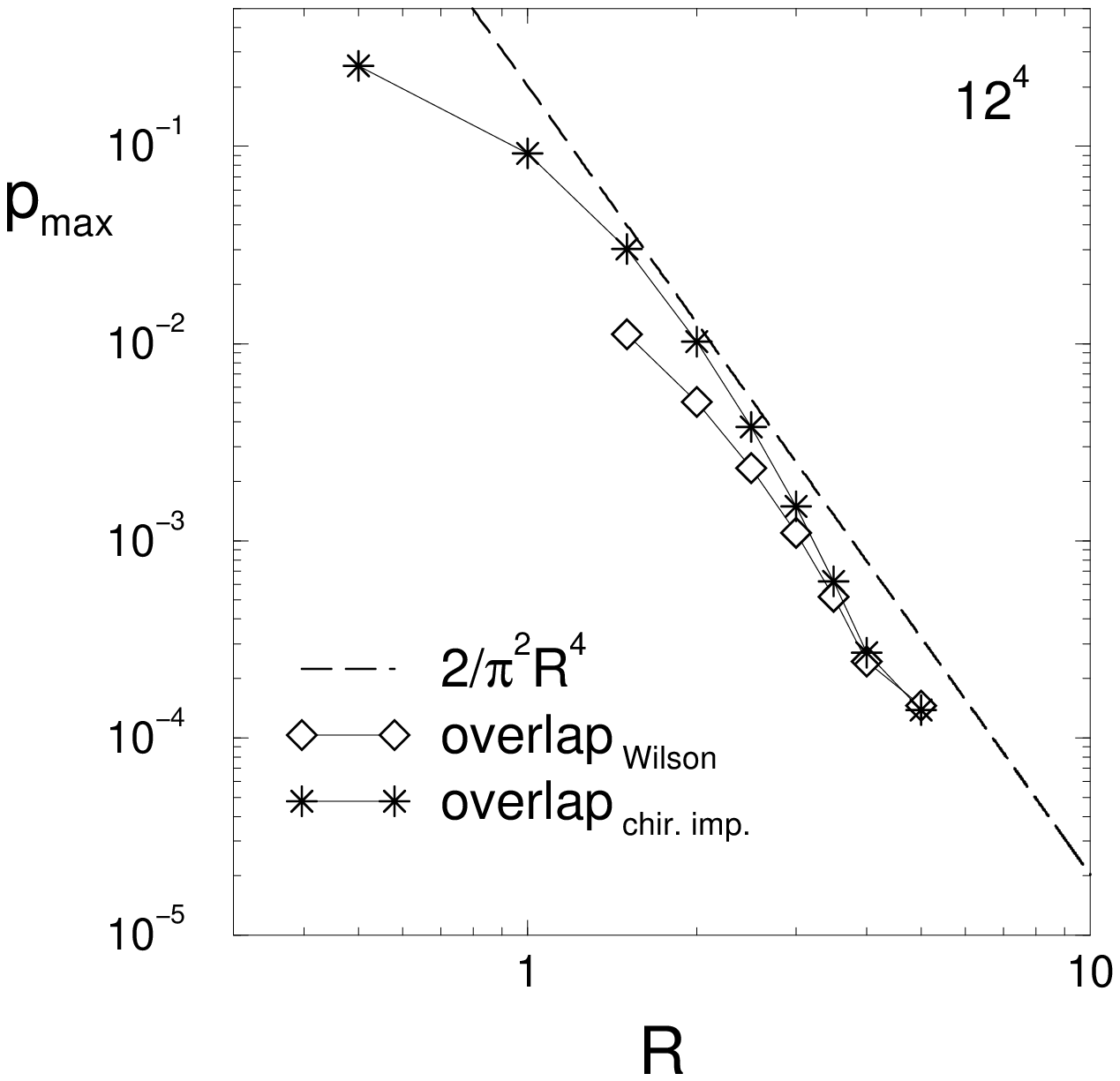,width=5.9cm}
\caption{The maximum $p_{max}$ of $p(x)$ as a function of 
the radius $R$ of the underlying instanton. In the plot on the left-hand side 
we show our results on $16^4$ lattices for
the Wilson overlap operator (diamonds), the chirally improved operator
(circles) and the Wilson operator (squares). The right-hand side plot 
provides a comparison of the Wilson overlap operator and the chirally 
improved overlap operator on $12^4$ lattices. 
\label{rhomax}}
\end{center}
\end{figure}

A second characteristic quantity of the zero mode is the maximum $p_{max}$ of 
$p(x)$, which in the continuum is given by
$p_{max} = p(0) = 2/(\pi^2 R^4)$. In Fig.~\ref{rhomax}
we plot our data for this quantity as a function of the instanton radius $R$.

Again the left-hand side plot shows the results for the chirally improved 
operator, the Wilson operator and the Wilson overlap operator, while the 
right-hand side compares the Wilson overlap operator with
the chirally improved overlap operator.
The overall picture is similar to the results for the inverse participation
ratio. The chirally improved operator gives the best results, while the 
Wilson overlap operator has again problems with reproducing 
the continuum results for smaller instantons. One finds 
that the amount of the deviation of the
Wilson overlap result from the continuum formula is already 50\% at $R = 2.5$ 
and increases further for smaller $R$. The situation is improved when using 
the chirally improved overlap operator.

We remark that we performed the same analysis with two changes 
of our setting: (1) Instead of using 1 as the center for the overlap 
projection we also used $1 + s$, with $s=0.1$, $s=0.2$ and $s=0.5$.
 Such an adjustment of the center of the
projection is known~\cite{jansenetal1} to optimize the localization 
properties of the Wilson overlap operator. We found that a variation of $s$ 
amounts to only small changes of $I$ and $p_{max}$ for the zero 
modes of the Wilson overlap operator.
(2) We also used a different discretization of the continuum instanton. 
We placed the instanton at the center of a hypercube. This allowed us
 to shrink the ``inner part'' of the gauge potential such that it
 only consisted of the interior of this hypercube, and we could then use
 the potential in the singular gauge on the whole lattice. 
 Also this modification does
not change the picture we obtained and shows that the results are
not very sensitive to the details of the discretization of the instanton.

To summarize, we find that the Wilson
overlap operator has significant problems 
with reproducing the continuum zero mode for instantons with $R \le 2.5$,
while the other operators, i.e.~the chirally improved operator, the 
Wilson Dirac operator and the chirally improved overlap operator 
do not show such a large deviation from the
continuum result.

  Why does the chirally improved overlap operator reproduce the
 continuum results better than the  Wilson overlap Dirac operator? 
 Possibly this is connected with the fact that the input Dirac 
 operator in the chirally improved case is already a much better
 approximation to a Ginsparg-Wilson operator than the Wilson 
 operator. 
 

For a typical simulation with $a \sim 0.1$~fm, our results for the 
Wilson overlap operator imply that structures smaller than $\sim 0.3$~fm 
will probably not be resolved properly. We expect that
the chirally improved operator with or without additional overlap projection 
fares better for such small structures.
\\
\\
{\bf Acknowledgements: } We would like to thank Tom DeGrand,
Peter Hasenfratz, Ivan Hip and Karl Jansen for
interesting discussions. This project was supported by 
the Austrian Academy of Sciences, the DFG and the BMBF. We thank 
the Leibniz Rechenzentrum in Munich for computer time on the Hitachi 
SR8000 and their operating team for training and support.


\begin{thebibliography}{1234567}
\newcommand{\bibi}[1]{\bibitem{#1}}
\newcommand{\authors}[1]{#1, }
\newcommand{\journal}[1]{#1}
\newcommand{\volume}[1]{#1}
\newcommand{\myyear}[1]{(#1)}
\newcommand{\page}[1]{#1}
\newcommand{\mytitle}[1]{}
\newcommand{\keywords}[1]{}
\newcommand{\kw}[1]{}


\bibi{GiWi82}
P.H.~Ginsparg and K.G.~Wilson, Phys.~Rev.~D~25 (1982) 2649.

\bibi{overlap}
R.~Narayanan and H.~Neuberger, Phys.~Lett.~B~302 (1993) 62,
Nucl.~Phys.~B~443 (1995) 305.

\bibi{fixpd}
P.~Hasenfratz, Nucl.~Phys.~B (Proc.~Suppl.) 63 (1998) 53;
P.~Hasenfratz, Nucl.~Phys.~B~525 (1998) 401;
P.~Hasenfratz, V.~Laliena and F.~Niedermayer,
Phys.~Lett.~B~427 (1998) 353;
P.~Hasenfratz, S.~Hauswirth, K.~Holland, Th.~J\"org,
F.~Niedermayer and U.~Wenger, 
Int.~J.~Mod.~Phys.~C 12 (2001) 691.

\bibi{domainwall}
P.M.~Vranas,
Nucl.\ Phys.\ Proc.\ Suppl.\   94  (2001) 177.

\bibi{Gaetal00}
C.~Gattringer, Phys.~Rev.~D 63 (2001) 114501;
C.~Gattringer and I.~Hip, Phys.~Lett. B~480 (2000) 112.

\bibi{Gaetal00b}
C.~Gattringer, I.~Hip and C.B.~Lang, Nucl.~Phys.~B597 (2001) 451.

\bibi{degrand}
T.~De Grand and A.~Hasenfratz, hep-lat/0103002.

\bibi{edwards}
R.G.~Edwards and U.M.~Heller, hep-lat/0105004.

\bibi{Gaetal01} 
C.~Gattringer, M.~G\"ockeler, P.E.L.~Rakow, S.~Schaefer and 
A.~Sch\"afer, hep-lat/0105023 (Nucl.~Phys.~B in print)
and hep-lat/0107016 (Nucl.~Phys.~B in print).

\bibi{domainwall2}
T.~Blum, N.~Christ, C.~Cristian, C.~Dawson, X.Liao,
G.~Liu, R.~Mawhinney, L.~Wu and Y.~Zhestkov, hep-lat/0105006.

\bibi{locchir} 
I.~Horv\'ath, N.~Isgur, J.~McCune and H.~B.~Thacker,
hep-lat/0102003.

\bibi{SchSh98}
T.~Sch\"afer and E.V.~Shuryak, Rev.~Mod.~Phys. 70 (1998) 323;
D.~Diakonov, Talk given at International School of Physics, 'Enrico Fermi', 
Course 80: Selected Topics in Nonperturbative QCD, Varenna, Italy, 
1995, hep-ph/9602375. 

\bibi{BaCa80} 
T.~Banks and A.~Casher, Nucl.~Phys.~B169 (1980) 103.

\bibi{wuppertal} 
I.~Hip, Th.~Lippert, H.~Neff, K.~Schilling
and W.~Schroers, hep-lat/0105001.

\bibi{DeGrand}
 T.~DeGrand and A.~Hasenfratz,
Phys.~Rev.~D 64 (2001) 034512.

\bibi{bietenholz} 
W.~Bietenholz,
Eur.~Phys.~J.~C 6 (1999) 537.

\bibi{jansenetal1}
P.~Hernandez, K.~Jansen and M.~L\"uscher,
Nucl.~Phys.~B 552 (1999) 363.

\bibi{jansenetal2}
P.~Hernandez, K.~Jansen and L.~Lellouch,
{\sl A numerical treatment of Neuberger's lattice Dirac operator},
in: Lecture Notes in Computational Science and Engineering 15,
A. Frommer, T. Lippert, B. Medeke, K. Schilling (Eds.),
Springer, Berlin 2000.

\bibi{clinst}
 I.A. Fox, M.L. Laursen, G. Schierholz, J.P. Gilchrist and M. G\"ockeler,
 Phys.\ Lett.\ B158 (1985) 332.

\bibi{arnoldi} 
D.C.~Sorensen, SIAM J.~Matrix Anal.~Appl.~13 (1992) 357; 
R.~B.~Lehoucq, D.C.~Sorensen and C.~Yang, ARPACK User's Guide, SIAM, New York,
1998.

\bibi{horvath} 
I.~Horv\'ath,
Phys.~Rev.~Lett.~81 (1998) 4063,
Phys.~Rev.~D 60 (1999) 034510;
W.~Bietenholz,
hep-lat/9901005.

\bibi{itoh} S.~Itoh, Y.~Iwasaki and T.~Yoshi\'e, Phys.~Rev.~D36 (1987) 527,
Phys.~Lett.~B 184 (1987) 375.

\end{thebibliography}
\end{document}